\documentclass{ws-ijmpe}

\begin{document}

\markboth{K. Sieja et. al.}{PARTICLE NUMBER CONSERVING APPROACH TO
  CORRELATIONS}

%
\catchline{}{}{}{}{}
%

\author{\footnotesize K. SIEJA$^{a,b}$, T.L. HA$^{c}$, P.QUENTIN$^{b}$
  and A. BARAN$^{a}$}

\title{PARTICLE NUMBER CONSERVING APPROACH TO CORRELATIONS}

\address{a) Institute of Physics, University of M. Curie-Sklodowska\\
  ul. Radziszewskiego 10, 20-031 Lublin, Poland\\
  b) Centre d'Etudes Nucl{\'e}aires de Bordeaux-Gradignan\\
  Chemin du Solarium, BP120, F-33175 Gradignan\\
  c) Department of Nuclear Physics, Vietnam National University Hano{\" i}\\
  334 Nguyen Tra{\" i}, Hano{\" i}, Vietnam}

\maketitle

\begin{history}
\received{October 19, 2006}
\end{history}

\begin{abstract}
  In the present work the so-called Higher Tamm-Dancoff Apporximation
  method is presented for the generalized case of isovector and
  isoscalar residual interactions treated simultaneously. The role of
  different particle-hole excitations and of proton-neutron pairing
  correlations in the ground
  state of the self-conjugate $^{64}$Ge nucleus is discussed.\\

\end{abstract}

\section{Introduction}
It was recognized already in the sixties that the proton-neutron
pairing may play a non-negligible role in the nuclei with equal
numbers of protons and neutrons due to the particulary large spatial
overlaps of the wave functions of both types of nucleons.  One of the
first methods developed to take into account such correlations were
BCS-type
approaches.\cite{Goswami64,Camiz65,Camiz66,Chen67,Goodman68,Goodman72}
It is however known that the BCS wave function violates the particle
number conservation which is especially harmful in the cases of weak
pairing and phase transitions.  A generic feature of the results of
BCS calculations generalized so as to accomodate neutron-proton
correlations ({\it i.e.}  that take into account isoscalar and
isovector pairs) in self-conjugate nuclei is a sharp transition from
like-particle to $T=0$ pairing modes.  Such a scenario is absent in
the particle number conserving approaches which allow for the
coexistence of different superfluid
phases.\cite{Satula00,Simkovic03,Sieja06}

In the present study we propose an approach which is free of the
deficiencies of the treatments {\`a} la Bogoliubov. This method,
called Higher Tamm-Dancoff Approximation (HTDA), was already applied
to study ground and isomeric states in heavier nuclei and to describe
pairing correlations in high spin states.\cite{Pillet02,Quentin04}
Here we extend the HTDA formalism to take into account the
proton-neutron pairing.

\section{The HTDA framework}
The purpose of the method is to describe the pair diffusion phenomenon
around the Fermi surface as a particular, yet very important, part of
some many quasiparticle excitations of the particle-hole type over a
Slater determinant vacuum $|\Psi_0\rangle$. Thus one can consider
dealing with a kind of higher Tamm-Dancoff approximation or
equivalently with a highly truncated shell-model calculation.  The
latter point of view makes it clear that the success ({\it i.e.,} the
fast convergence in terms of the number of considered quasiparticles)
of such a truncation scheme will depend on the realistic character of
the vacuum in use. Such a favorable situation may be expected when the
vacuum is defined self-consistently from the mean-field carrying most
of the single-particle (sp) properties (even those yielded by the
correlations) associated with a given effective Hamiltonian
\begin{equation}
\hat H=\hat K+\hat V\,,
\label{htot}
\end{equation} 
where $\hat K$ is the kinetic energy and $\hat V$ an effective
interaction.  It is advisable to choose for the wave function
$|\Psi_0\rangle$ the Hartree-Fock solution coming from the HF
potential $\hat V_{\rm HF}$ self-consistently obtained from the
many-body reduced density matrix $\hat\rho$ of the correlated solution
for the desired number of particles and possibly taking into account
various constraints ({\it e.g.}, deformation) and symmetries ({\it
  e.g.}, time-reversal symmetry).  We have thus
\begin{equation}
\hat H_{\rm HF}|\Psi_0\rangle=E_0 |\Psi_0\rangle\,,
\end{equation}
where 
\begin{equation}
\hat H_{\rm HF}=\hat K+\hat V_{\rm HF}\,,
\end{equation}
$\hat V_{\rm HF}$ being the one-body reduction of $\hat V$ for $|\Psi_0\rangle$.

The $|\Psi_0\rangle$ vacuum may now serve to construct an orthonormal
N-body basis in which we will diagonalize Hamiltonian (\ref{htot}). In
principle to build this basis we should consider the Slater
determinant corresponding to the 0p0h state and all the particle-hole
excitations: 1p1h $|\Psi_1\rangle$, 2p2h $|\Psi_2\rangle$, 3p3h
$|\Psi_3\rangle$ and so on. It is however obvious that in practice we
need to truncate the space of particle-hole excitations. Based on the
former studies in the HTDA framework\cite{Pillet02} we may assume that
the components of the pair excitation type will be dominant in the
ground state. In the section \ref{results} we will show for the case
of $pp$ and $nn$ correlations that the other components are indeed of
a minor importance.

The total ground state wave function can be decomposed in the
following way ($\tau$ and $\tau'$ denoting two different charge
states)
\begin{eqnarray}
  |\Psi\rangle&\equiv&|\Psi^\tau\otimes\Psi^{\tau'}\rangle=\chi_{00}|\Psi_0^\tau\otimes\Psi_0^{\tau'}\rangle
  +\sum_{(1{\rm p}1{\rm h})_\tau}\chi_{10}|\Psi_1^\tau\otimes\Psi_0^{\tau'}\rangle
  +\sum_{(1{\rm p}1{\rm h})_{\tau'}}\chi_{01}|\Psi_0^\tau\otimes\Psi_1^{\tau'}\rangle\nonumber\\
  &+&\sum_{(1{\rm p}1{\rm h})_{\tau}(1{\rm p}1{\rm h})_{\tau'}}\chi_{11}|\Psi_1^\tau\otimes\Psi_1^{\tau'}\rangle
  +\sum_{(2{\rm p}2{\rm h})_\tau}\chi_{20}|\Psi_2^\tau\otimes\Psi_0^{\tau'}\rangle+
  \sum_{(2{\rm p}2{\rm h})_{\tau'}}\chi_{02}|\Psi_0^\tau\otimes\Psi_2^{\tau'}\rangle\nonumber\\
  &+&\cdots
\label{mbstpn}
\end{eqnarray}

The ensemble of products of Slater determinants
$|\Psi_i^\tau\otimes\Psi_j^{\tau'}\rangle$ represents a complete
orthogonal basis of the N neutrons and Z protons physical space with
real coefficients $\chi_{m}$ fulfilling the relation
\begin{equation}
\sum_{m} \chi_{m}^2=1
\end{equation}
which guarantees the normalization of the function (\ref{mbstpn}). It
is clear that this function has a good particle number $\langle
\Psi|\hat N|\Psi\rangle=A$.  To obtain the correlated ground state one
should perform the diagonalization of the Hamiltonian (\ref{htot}) in
the given many-body basis.

Let us rewrite the Hamiltonian (\ref{htot}) in the following form
\begin{eqnarray}
\hat H&=&\hat K+\hat V_{\rm HF}-\langle\Psi_0|\hat V|\Psi_0\rangle+\hat V-\hat V_{\rm HF}
+\langle\Psi_0|\hat V|\Psi_0\rangle\nonumber\\
&=& \langle\Psi_0|\hat H|\Psi_0\rangle+\hat H_{\rm IQP}+\hat V_{\rm res}\,,
\label{h2}
\end{eqnarray}
where the independent quasiparticle Hamiltonian $\hat H_{\rm IQP}$ reads
\begin{equation}
\hat H_{\rm IQP}=\sum_{i}\xi_i \eta_{i}^\dagger\eta_{i}\,,
\end{equation} 
where $\eta_{i}^{\dagger}$ is the particle creation operator
$a_{i}^\dagger$ for $i$ being a particle (unoccupied) state and to the
anihhilation operator $a_{i}$ in the case of hole (occupied) states.
For $\xi_{i}$ we have: $\xi_{i}=e^{i}$ or $\xi_{i}=-e^{i}$, with
$e^{i}$ indicating the single-particle (HF) energy corresponding to
the particle or hole case, respectively.  As seen from Eq. (\ref{h2}),
the residual interaction reads
\begin{equation} 
\hat V_{\rm res}=\hat V-\hat V_{\rm HF}+\langle\Psi_0|\hat V|\Psi_0\rangle\,.
\label{vres}
\end{equation}

The matrix element of the above given Hamiltonian in the
multi-particle multi-hole basis takes the form
\begin{eqnarray}
  H_{ij}&=&\left(\langle\Psi_0|\hat H|\Psi_0\rangle+
    \sum_{\tau=p,n}E_{\rm p-h}^{i\tau}\right)\delta_{ij}+\langle\Psi_i|\hat V_{\rm res}|\Psi_j\rangle\,,
\label{hijkl}
\end{eqnarray}
where $\sum_{\tau=p,n}E_{\rm p-h}^{i\tau}$ is the total particle-hole
excitation energy calculated with respect to the vacuum
$|\Psi_0\rangle\equiv|\Psi_0^\tau\otimes\Psi_0^{\tau'}\rangle$.  The
residual interaction comprising only one- and two-body operators will
merely contribute to the matrix elements in cases where the two
many-body states are identical or differ by one or two particles.  The
many-body matrix elements of $\hat V_{\rm res}$ can be evaluated with
the use of the Wick theorem (for details see Ref.\cite{Long}) and
expressed as sums of proper two-body matrix elements of the
interaction $\hat V$.

In actual calculations with our approach we have used the Skyrme force
for the HF part of the problem. Since the majority of the Skyrme
forces is not well suited to reproduce the data in the
particle-particle channel we apply a $\delta$-force to define the
residual interaction.  Using the zero-range force is well established
in the context of like-particle and-- to a lesser extent-- of the
proton-neutron pairing. To assess the validity of our approach at this
early stage of the investigation of the proton-neutron correlations
within the HTDA framework a substitution
\begin{equation}
  \hat V\Rightarrow \hat V_\delta=\sum_{T=0,1}V_{0}(T)\delta(\vec r_{12})\Pi^S\Pi^T\, 
\end{equation} 
is therefore done for the residual interaction.  The operators
$\Pi^S\Pi^T$ project onto spin-isospin subspaces
\begin{eqnarray}
\hat\Pi^S&=&\frac{1}{2}(1-(-1)^S P^\sigma)\,,\\
\hat\Pi^T&=&\frac{1}{2}(1-(-1)^T P^\tau)\,,  
\label{pispit}  
\end{eqnarray}
where $P^\sigma={1/2}(1-\vec \sigma_1\cdot\vec\sigma_2)$,
$P^\tau={1/2}(1-\vec \tau_1\cdot\vec\tau_2)$ are the standard spin and
isospin exchange operators.

\section{Results}
\label{results}
The main challenge in the exact pairing diagonalizations methods and
shell-model like calculations is the large dimension of the matrix to
be diagonalized. In the present study we use the Lancz{\"o}s
algorithm\cite{Lanczos50} within the code developed by Parlett and
Scott\cite{Parlett79} which allows to search for the lowest energy
solutions in a reasonable time. However, as will be shown in the
following, the advantage of the HTDA method is that the amount of the
many-body configurations within a given single-particle space that
need to be taken into account is quite limited and the matrix
diagonalization problem is far from attaining the difficulty level of
typical shell-model calculations.

The calculations within the HTDA framework which we will discuss here
are performed for the ground state of the $^{64}$Ge nucleus as
obtained in a Skyrme-HF+BCS approach.  We use the SIII force for the
particle-hole channel, seniority force in the BCS part and volume
$\delta$ interaction to calculate two-body matrix elements in the HTDA
part.  The HTDA results presented here are non self-consistent.
Indeed, only one diagonalization of the HTDA Hamiltonian matrix has
been carried out for a Hamiltonian $\hat H_{\rm HF}$ obtained with the
density $\hat\rho$ corresponding to the HF+BCS calculation.

\subsection{Limiting case: pp and nn interactions}
In the limiting case of vanishing proton-neutron correlations we
obtain two decoupled problems, one for each charge state. Let us
discuss first the aspects of the basis truncation, both in
single-particle and many-particle many-hole excitations terms. If the
assumptions of the BCS theory are to hold, we may expect that the most
important part of the ground state correlations will correspond to the
pair excitations, {\it i.e.} the situation where two particles in
Kramers degenerate hole orbitals are promoted to two Kramers
degenerate particle levels. In what follows we will restrict the
many-body states first to take into account all possible 1p1h and 2p2h
excitations including thus the one pair excitations in addition to the
particle-hole vacuum. In further calculations only the states
corresponding to one, two and three pairs excitations will be
considered.

\begin{table}[t]
  \caption{Numbers of many-body configurations of different types
    depending on the number of sp levels in the window.}
\label{tab-1}
\begin{center}
\begin{tabular}{cccccc}
  \hline
  \hline
  Number of sp levels & 1p1h &1 pair& 2p2h & 2pairs&3 pairs\\ 
  \hline
  &&&&&\\
  12  & 10 & 40 & 260  & 280   & 560\\
  20  & 22 & 96 & 1744 & 1848  & 12157\\
  28  & 48 & 180& 6596 & 6758  & 58404\\  
  36  & 76& 260 &13706& 14230  &$>$100000\\  
  &&&&&\\
  \hline
  \hline
\end{tabular}
\end{center}
\end{table}

Having chosen a configuration space size in terms of the complexity of
many-particle many-hole states, one has to further truncate on
single-particle levels from which these configurations will be built.
We are facing thus a situation met in customary BCS calculations.
Typically, we will limit our sp subspace to the so-called
configuration space window defined as
\begin{equation}
e_F\pm E_{\rm cut}^{\rm sp}  \,,     
\end{equation}
where $e_F$ is the Fermi energy (defined as the average between the
single-particle energies of the last occupied and first unoccupied
levels) and $E_{\rm cut}^{\rm sp}$ the cut-off energy.  The actual
value of the latter should be chosen such that the inclusion of
additional sp levels does not introduce any significant physical
consequences but only a possible slight renormalization of the
relevant quantities.  In Table \ref{tab-1}, to exemplify the matrix
dimensions, we give numbers of configurations corresponding to
different particle-hole excitations depending on the number of sp
levels included in the configuration space window.

Since in the HTDA calculations we are using a $\delta$-force it is
clear that both the coupling constant and the cut-off parameter are
necessary to define fully the interaction. It is then indispensable to
fix the $\delta$-force strength in our approach to reproduce physical
quantities, {\it e.g.}, the phenomenological 3-point gaps for the
considered nucleus ($\Delta_n$=1.48 MeV, $\Delta_p$=1.14 MeV evaluated
with experimental masses of Ref.\cite{Audi03}). Assuming that the
appearance of the pairing gap is related to a breaking of the Cooper
pair of the lowest energy, we perform the HTDA calculations with one
level (neutron or proton) closest to the Fermi energy blocked and
adopt the difference of the expectation value of $\hat V_{\rm res}$ in
the two calculations as a proper measure of the pairing correlations
that can be compared to the experimental odd-even mass differences.
Namely, we define
\begin{equation}
\Delta=[E(n)-E_{\rm IQP}(n)]-[E(n-1)-E_{\rm IQP}(n-1)]\,,
\end{equation}
where $E=\langle\Psi|\hat H|\Psi\rangle$, $E_{\rm
  IQP}=\langle\Psi|\hat H_{\rm IQP}|\Psi\rangle$ and $n$ indicates the
number of sp levels in the configuartion space window. Since we deal
with the $N=Z$ nucleus one value of the coupling constant $V_0$ and of
the cut-off energy $E_{\rm cut}^{sp}=10$MeV is adopted for both charge
states in further calculations.

\begin{table}[t]
  \caption{Percentage of different components of the ground state correlated wave function. The results concern 
    the case where all types of particle-hole excitations up to the 2$^{\rm nd}$ order are taken into account.}
\label{tab-2}
\begin{center}
\begin{tabular}{lcccc}
\hline
\hline
         & 0p0h & 1p1h & 1 pair & 2p2h (all included)\\
\hline
&&&&\\
neutrons & 69.6 & 0.003 & 29.6 & 30.4 \\
protons  & 70.6 & 0.005 & 29.3 & 28.3 \\
&&&&\\
\hline
\hline
\end{tabular}
\end{center}
\end{table}

First, let us discuss the result of the diagonalization of the
Hamiltonian matrix in the many body basis containing all 1p1h and 2p2h
components. The decomposition of the correlated function for neutrons
and protons is given in Table \ref{tab-2}.  As expected, the main
contribution beyond the 0p0h vacuum comes from one pair excitations.
The 1p1h part is fully negligible while the 2p2h excitations of other
types constitute less than 1\% of the correlated wave function so
restricting the many-particle many-hole excitations to study only pair
excitations in addition to the vacuum seems well justified.  Of course
it allows for a great simplification from the calculational point of
view, as the number of pair excitations is very modest as compared to
the number of all possible particle-hole excitations (cf. Tab.
\ref{tab-1}).

The same procedure is employed in the case of calculations where one,
two and three pairs excitations in addition to the vacuum are
considered. In Table \ref{tab-3} the decomposition of the correlated
wave function is given for neutrons and protons in three cases: when
only 1 pair excitations are considered in addition to the vacuum
state, when 2 pairs are added and finally, when 3 pairs are as well
embedded in the calculations. We also give the corresponding values of
the correlation energy defined as the difference of the mean values of
the Hamilton operator of the system in correlated and uncorrelated
states
\begin{equation}
E_{\rm corr}=\langle\Psi|\hat H|\Psi\rangle-\langle\Psi_0|\hat H|\Psi_0\rangle\,.
\end{equation}
As seen, one pair excitations contribute the most in all cases, two
pairs excitations give only about 2\% of the total amount while the
probability for three pairs is negligible. Including the latter has a
minimal influence on both the correlation energy and on the percentage
of other components.  However, the presence of two pairs boosts the
population of the one pair excitation content of the correlated wave
function, as already observed in the calculations of Ref.\cite{Long}.
This entails that to describe accurately the one pair excitation
phenomenon two pair transfers are a priori needed.  Nevertheless, at
this early stage of our study of proton-neutron pairing correlations
we will exclude 2 pairs components for obvious practical reasons.
\begin{table}[t]
\caption{Percentage of various components of the correlated wave function
and the correlation energy values
for neutrons and protons. Results of three calculations are reported:
with 1 pair excitations added to the vacuum,
with 1 and 2 pairs and with 1, 2 and 3 pairs included.}
\label{tab-3}
\begin{center}
\begin{tabular}{lccccc}
  \hline
  \hline
  & 0p0h & 1 pair & 2 pairs &3 pairs & $E_{\rm corr}$ (MeV)\\
  \hline
  &&&&&\\
  neutrons & 65.4 & 34.6 & --  & -- & -1.16\\
  & 49.55& 48.5 & 1.95& -- & -1.43\\
  & 49.0 & 48.9 & 2.1 &0.02& -1.45\\
  &&&&&\\	 
  \hline
  &&&&&\\

  protons  & 65.3 & 34.7 & -- & -- & -1.18 \\
  & 47.7 & 50.2 & 2.1& -- & -1.48 \\
  & 46.9 & 50.7 & 2.3 & 0.04 & -1.50\\ 
  &&&&&\\
  \hline
  \hline
\end{tabular}
\end{center}
\end{table}

\subsection{General case: $T=1$ and $T=0$ pairing interaction}
As we already stated we aim at including the $T_{z}=0$ pairing
correlations within the HTDA framework.  As above discussed in this
preliminary study we limit the configuration space to the so-called
one pair excitations in all channels.

\begin{figure}
\begin{center}
\includegraphics[scale=0.9]{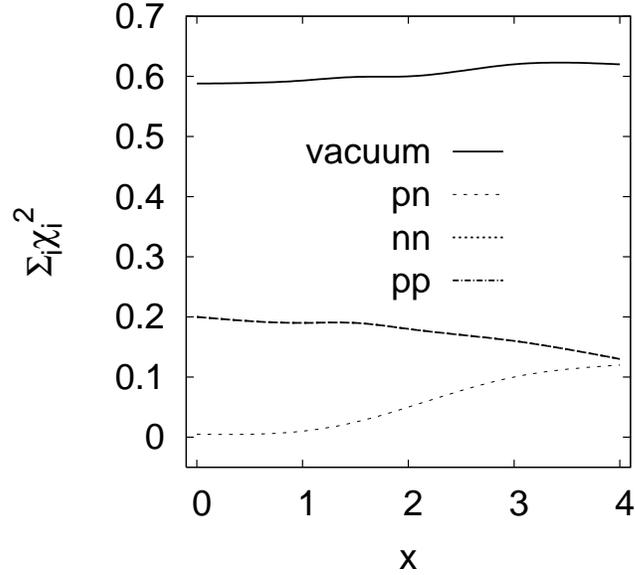}
\end{center}
\caption{Total probability of different type of components
  (vacuum-(0p0h)$_p\otimes$(0p0h)$_n$, pn-(1p1h)$_p\otimes$(1p1h)$_n$,
  pp-(0p0h)$_n\otimes$(2p2h)$_p$, nn-(2p2h)$_n\otimes$(0p0h)$_n$) in
  the correlated wave function $|\Psi\rangle$ versus the ratio $x$ of
  the $\delta$-force strengths in isoscalar $(T=0)$ and isovector
  $(T=1)$ channels.}
\label{fig-1}
\end{figure}

Since proton-neutron pairs may exist in both $T=0$ and $T=1$ channels
the problem of adjusting the strength of the isoscalar pairing arises
here. The interplay of $T=0$ and $T=1$ pairing being rather unclear,
we will adopt the same value of the isovector pairing strength for
$pp$, $nn$ and $pn$ channels and will treat the $T=0$ interaction
strength as a free parameter.\footnote{A considerable amount of $pn$
  pairing correlations can be as well obtained in an isospin broken
  model with $V_0^{\tau\tau'}>V_0^{\tau\tau}$, nevertheless we do not
  see any argument supporting such a choice.}

\begin{figure}
\begin{center}
\includegraphics[scale=0.9]{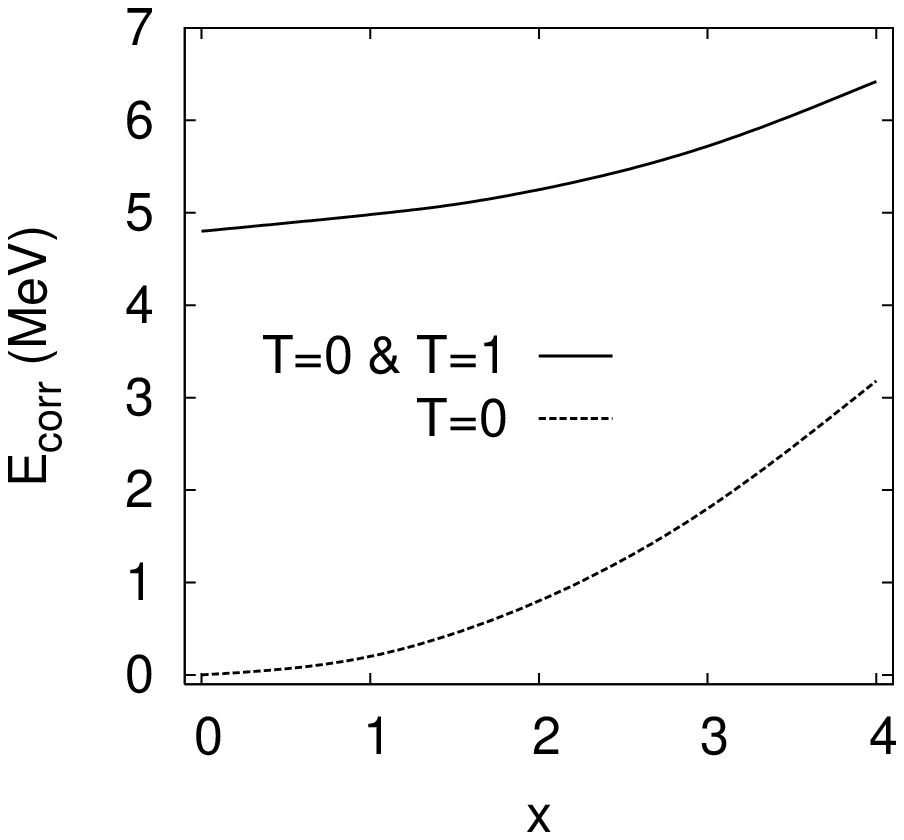}
\end{center}
\caption{Absolute value of the correlation energy plotted against the
  ratio $x$ of the interaction strengths in isoscalar and isovector
  channels. The results concern calculations with $T=0$ $pn$ pairs
  only and with all possible pairs taken into account.}
\label{fig-2}
\end{figure}

Within a given configuration space and $T=1$ strength adjusted as
described in the precedent we perform the calculations as a function
of the increasing $T=0$ coupling constant. In Fig. \ref{fig-1} the
evolution of the probability of different components is plotted
against the ratio
\begin{equation}
x=V_0(T=0)/V_0(T=1)\,. 
\end{equation}
The probability of the $T_z=0$ pair excitation becomes siginificant
with respect to that of $pp$ or $nn$ pair excitation for $x>2$ which
corresponds to the ratio value at which a $pn$ superfluid solution is
found in BCS($\delta$)-type of calculations in this
nucleus.\cite{Sieja06} The gain in energy due to the $pn$ coupling is
non-negligible which can be seen from Fig. \ref{fig-2} where the
contribution of different components to the correlation energy is
depicted. It is worth noting that the increase of the energy is of a
similar magnitude like that of BCS calculations with the approximate
particle number projection of the Lipkin-Nogami (LN)\cite{Sieja06}
nonetheless a certain gain in the HTDA correlation energy due to the
$pn$ coupling occurs already in the range of parameters for which no
$pn$ pairing solution is found in the BCS+LN method ($x\le2$).

\section{Summary}
We have presented a method explicitely conserving particle number in
the context of describing isovector and isoscalar pairing
correlations.  The inclusion of $T=0$ pairs leads to a constant
increase of the absolute value of the correlation energy. The expected
benefit of the HTDA framework in yielding a much smoother decrease of
correlations than when using a Bogoliubov quasiparticle vacuum in weak
pairing regimes is confirmed.

\section{Acknowledgements}
One of the authors (K.S.) is indebted to Jacek Dobaczewski for
fruitful discussions.  This work is partly supported by the Committee
of Scientific Research (KBN) of the Polish Ministry of Science and
Education under contract No.  1P03B13028 (2005-2006) and the French
Embassy in Poland.

\end{document}